\documentclass[11pt,twoside]{article}

\usepackage{asp2014}

\aspSuppressVolSlug
\resetcounters

\begin{document}

\title{Calibration of ground based survey data using Gaia: Application to DES}


\author{Koshy George,$^1$ Thomas Vassallo,$^1$ Joseph Mohr,$^1$ Mohammad Mirkazemi,$^1$ Holger Israel,$^1$ Jelte T. A. de Jong,$^2$ Gijs A. Verdoes Kleijn$^2$}

\affil{$^1$Faculty of Physics,Ludwig-Maximilians-Universit{\"a}t, Scheinerstr. 1, Munich, 81679, Germany ;\email{Koshy.George@physik.lmu.de}}

\affil{$^2$Kapteyn Astronomical Institute, University of Groningen, P.O. Box 800, Groningen, 9700 AV, the Netherlands}

\begin{abstract}
The calibration of ground based optical imaging data to photometric accuracy of 10 mmag over the full survey area and to color uniformity better than 5 mmag on the scale of the VIS focal plane is a key science requirement for the Euclid mission. These accuracies enable stable photometric redshifts of galaxies and modeling of the color dependent VIS PSF for weak lensing studies. We use the Gaia photometry to calibrate the $g/r/i/z$ magnitudes of Dark Energy Survey (DES) data to meet the stringent Euclid requirements. The Gaia G band magnitude along with the BP-RP color information of stars observed in the DES single epoch (SE) exposures are used to derive the transformation from Gaia to DES photometry for individual CCDs and to characterize persistent photometric errors across the DECam focal plane. We use the color dependence of these persistent errors to characterize the $g/r/i/z$ bandpass variations across the DECam focal plane.

\end{abstract}



\section{Introduction}

The uniform, high quality Gaia stellar photometry provides a stable photometric reference frame  over the whole sky. The density of Gaia calibrators is high enough that individual CCD images from ground based surveys can be calibrated to $\sim$ 1mmag precision. Ongoing testing of Gaia DR2 suggests that the systematic floor in this calibration is currently at comparable or even lower levels than one can achieve with self-calibration techniques in surveys like DES. In the following sections we demonstrate a mapping from the Gaia photometry to the DES $g/r/i/z$ bands and present first results of a test for bandpass variations in the DECam focal plane.


\section{Gaia transformed to DES}

We used the Gaia DR2 \citep{2018A&A...616A...4E} and DES DR1 \citep{2018ApJS..239...18A} photometry data of stars to create a transformation function between the broad band Gaia G band filter and the relatively narrow band DES $(g/r/i/z)$ filter system using Gaia G, BP, RP and independently calibrated DES photometry of stars covering the Euclid Deep Field South (EDFS). EDFS covers an RA range of 55.10$^{\circ}$ to 67.40$^{\circ}$ and a DEC range of -51.90$^{\circ}$ to -45.10$^{\circ}$ and has been chosen for a Euclid science survey. We used the DES DR1 data from the EDFS field with 11,500 stars overlapping with Gaia DR2 data to construct the transformation function. A polynomial function ($T_{func}$) of order 10 is found to be a good fit for the Gaia to DES $g/r/i/z$ filter transformation as shown in the equation below and in Figure \ref{fig1}.

\begin{displaymath}
DESmag(g/r/i/z) = Gaia\ Gmag + {T_{func}}(BP-RP)
\end{displaymath}

\articlefigure[width=0.5\textwidth]{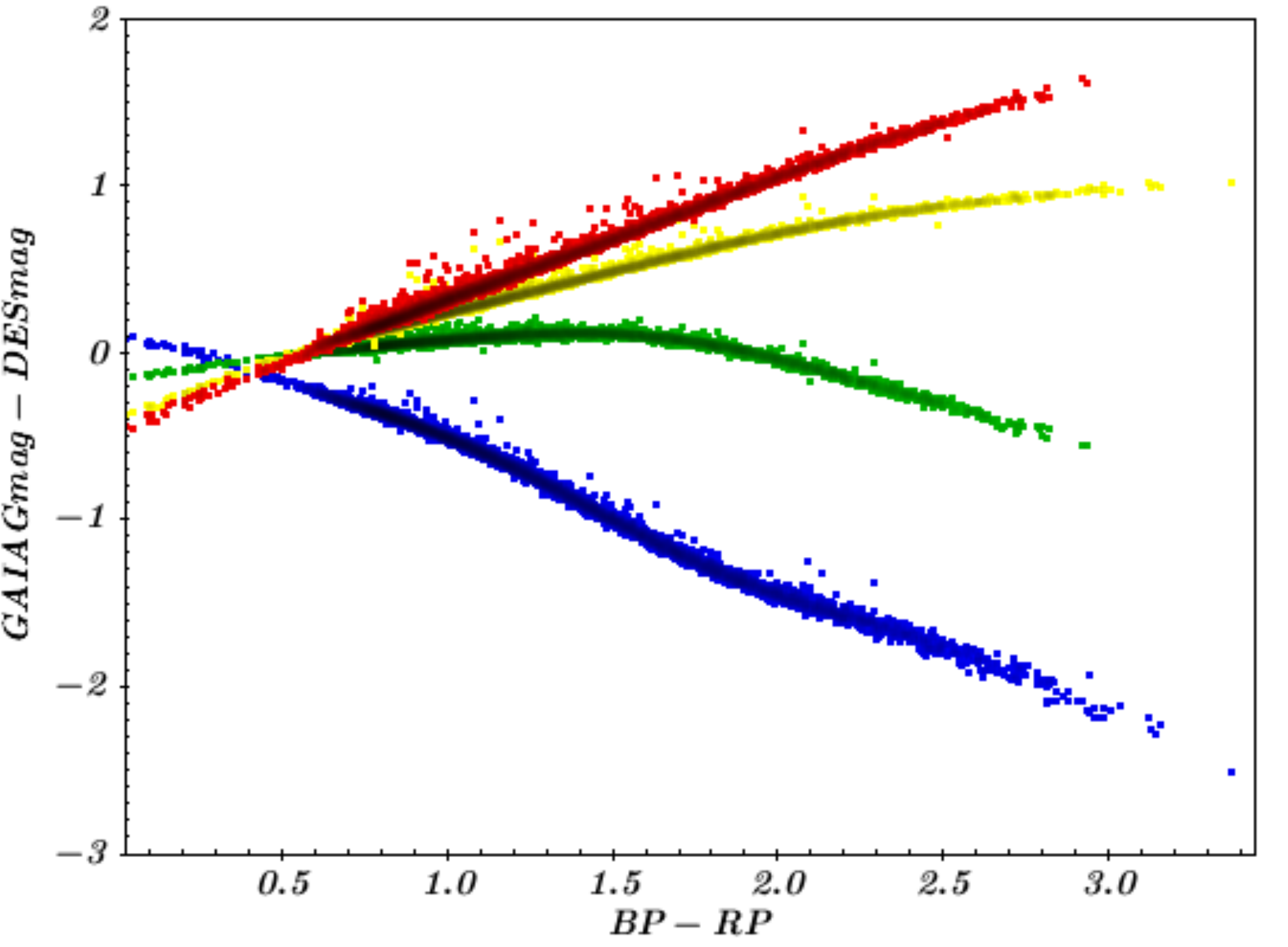}{fig1}{The transformation between Gaia G and DES $g/r/i/z$ magnitudes as a function of Gaia BP-RP color can be fit with a polynomial function of order 10. Points are color coded with DES g (blue), r (green), i (yellow) and z (red) with corresponding NMAD scatter about the transformation function of 1.3, 0.9, 0.9 and 1.4 mmag, respectively.}

\section{Zeropoint computation for DES single epoch images}

\articlefigure[width=0.5\textwidth]{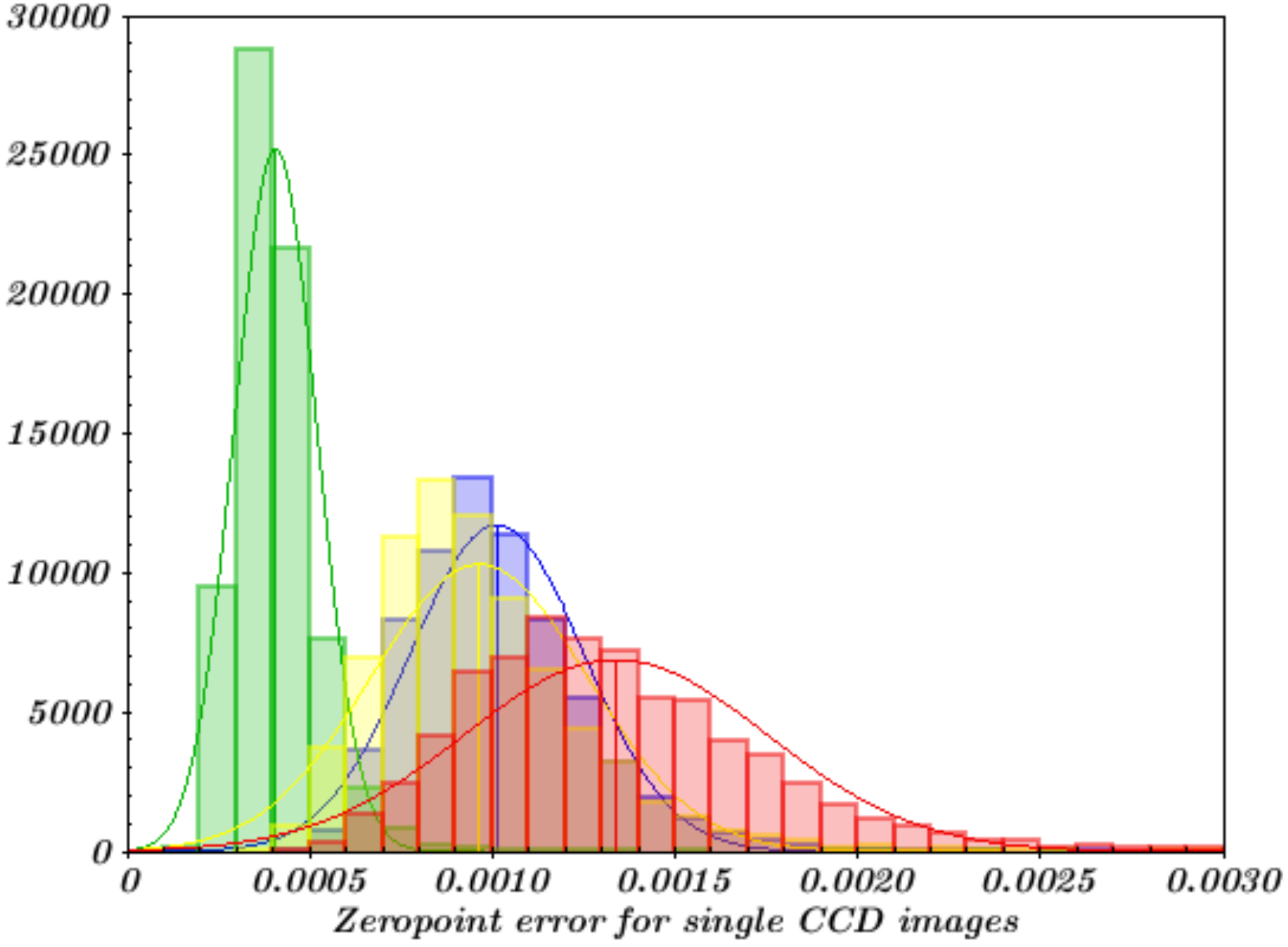}{fig2}{The distribution of Gaia calibrated zero point uncertainties (statistical only, using NMAD scatter presented in Figure \ref{fig1}) for SE catalogs. Bins are color coded for each DES band as in Figure \ref{fig1}.}

Our primary goal is the calibration of individual CCD images (SE images) in DES $(g/r/i/z)$ using the associated catalogs and the Gaia data. We can achieve this by computing the zeropoint (zp) for SE images using the transformation function and the Gaia G, BP, RP and DES $(g/r/i/z)$ instrumental magnitude (DESim) of stars in SE catalogs as shown below.

\begin{displaymath}
zp\ =\ median[DESim(g/r/i/z) - Gaia\ Gmag - T_{func}(BP-RP)]_{N stars}
\end{displaymath}

The characteristic statistical uncertainty in the zp is the scatter about the transformation function divided by $\sqrt{N}$, where N is the number of stars. The uncertainty distributions for zero points computed for DES $(g/r/i/z)$ SE images in the EDFS field are shown in Figure \ref{fig2}. We are effectively calibrating DES $g/r/i/z$ data using Gaia photometry at $\sim$ 1 mmag level uncertainty. Current estimates of the systematic floor in the Gaia photometry is 1.8 mmag.


\section{Bandpass variation across DECam focal plane}

The DECam focal plane consists of 62 2K $\times$ 4K pixel CCDs arranged in a hexagonal pattern. The transformation functions between Gaia and DES provide an opportunity to predict the DES $(g/r/i/z)$ magnitude of stars lying in any region of the CCD in the DECam focal plane using the Gaia DR2 G, BP, RP magnitude information of stars. The DES $(g/r/i/z)$ filter bandpass can show spatial variations across the focal plane, which we can quantify using the transformation function. Any deviation from the predicted DES $(g/r/i/z)$ magnitude with respect to the observed magnitude can be an indication of the change in transformation function due to band pass variation. We quantify the deviation between predicted and observed DES $(g/r/i/z)$ magnitudes of stars in all SE catalogs for a given CCD. This deviation is found to vary with the BP-RP color of stars for each CCD. An example of the variation in the $i-$band across a single CCD subregion is shown in Figure \ref{fig3}. The best fit relation is shown in red and the slope of relation (color term) is an indication for the DES filter bandpass variation. We divided the CCD into four subregions and the focal plane map of color term variations in the DES $(g/r/i/z)$ filters are shown in Figure \ref{fig4}. There exist statistically significant color term variations (and therefore bandpass variations) across the focal plane for the DES $(g/r/i/z)$ filters.

\articlefigure[width=0.5\textwidth]{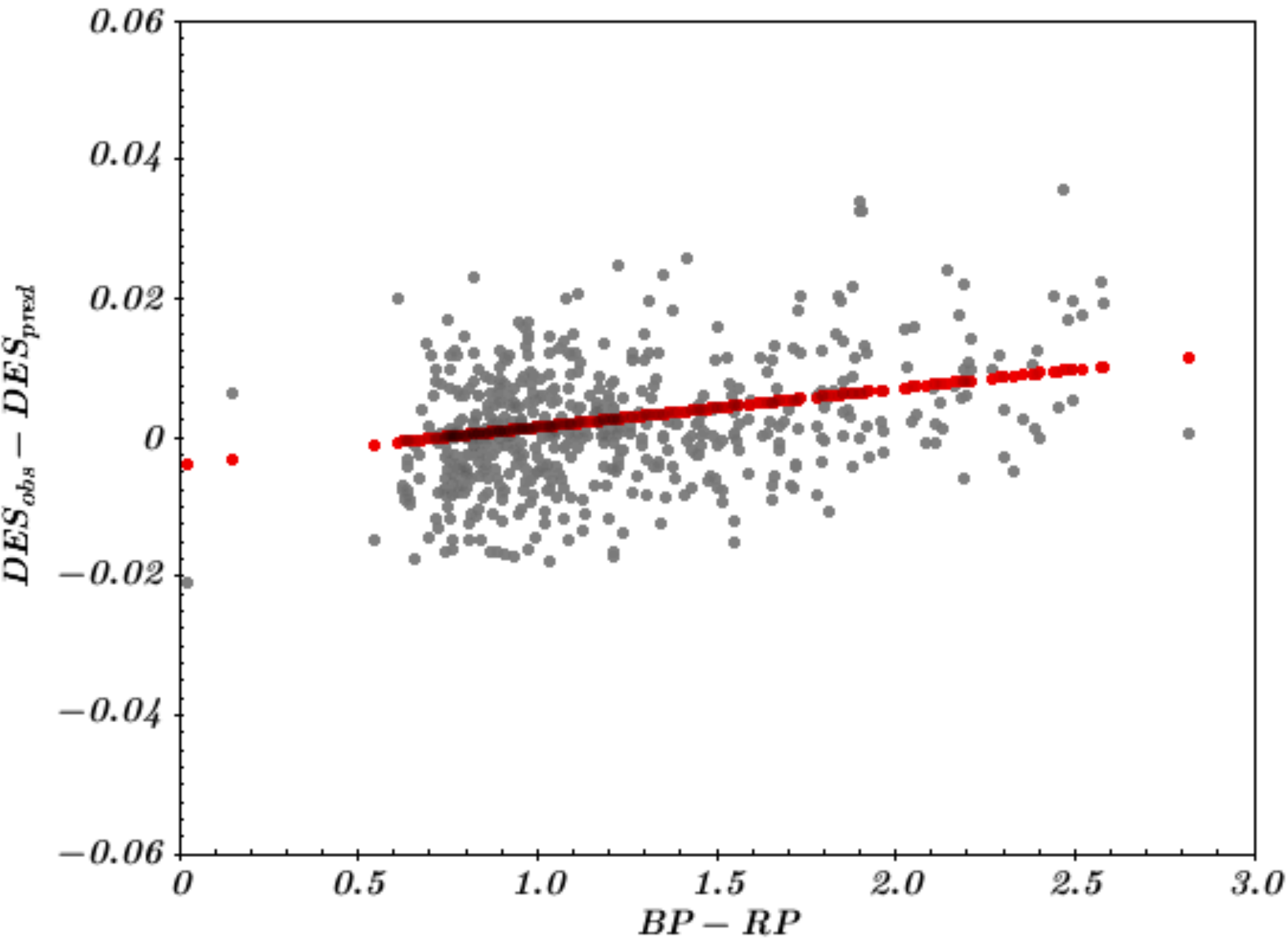}{fig3}{The deviation of DES observed and predicted magnitude of stars from SE catalogs in a subregion for a single CCD is plotted against the Gaia BP-RP color. Any color dependence in this deviation provides evidence for a bandpass difference between this CCD region and the average transformation function shown in Figure 1. The best fit relation is
 shown in red.}

\articlefigurefour{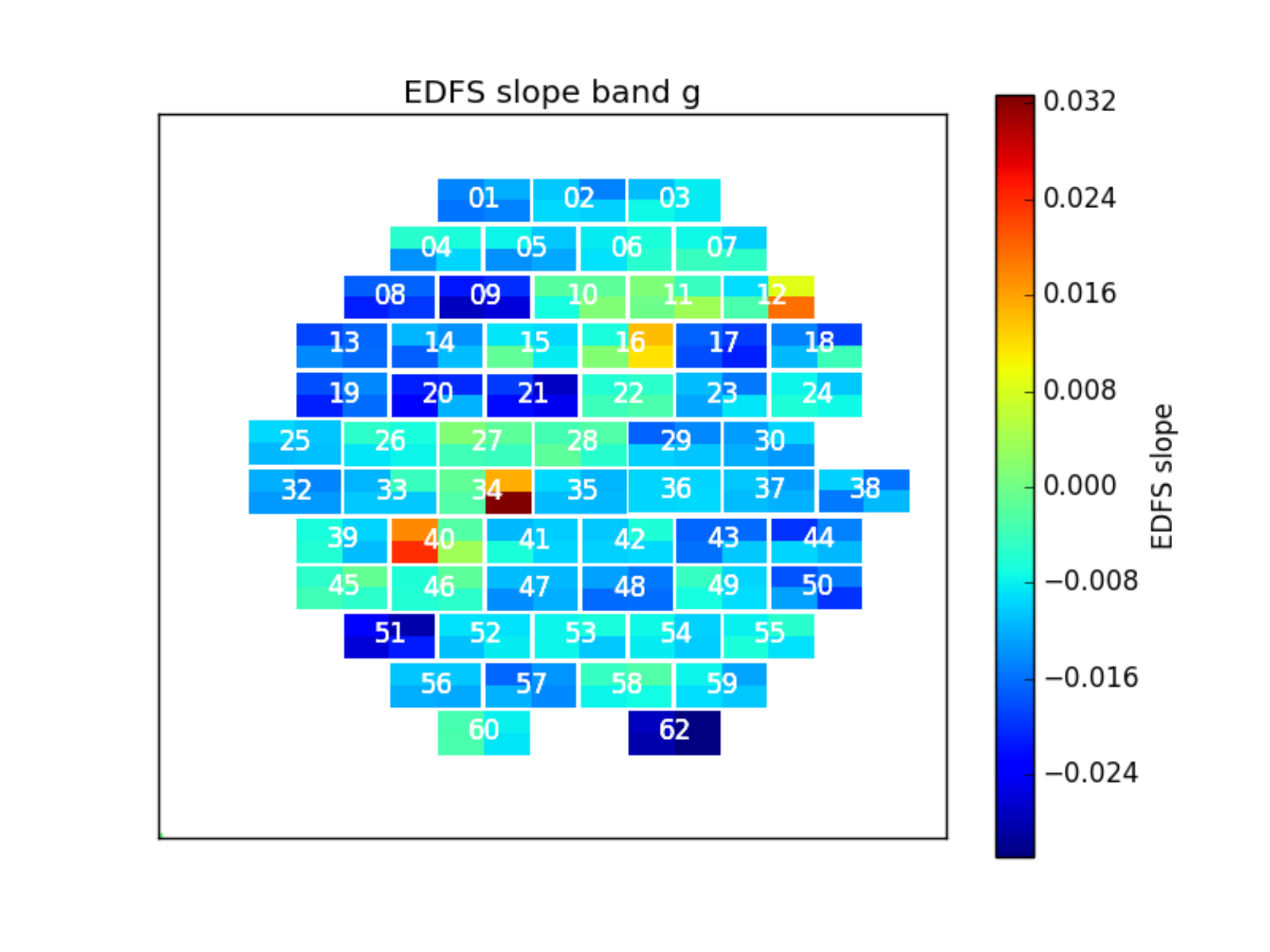}{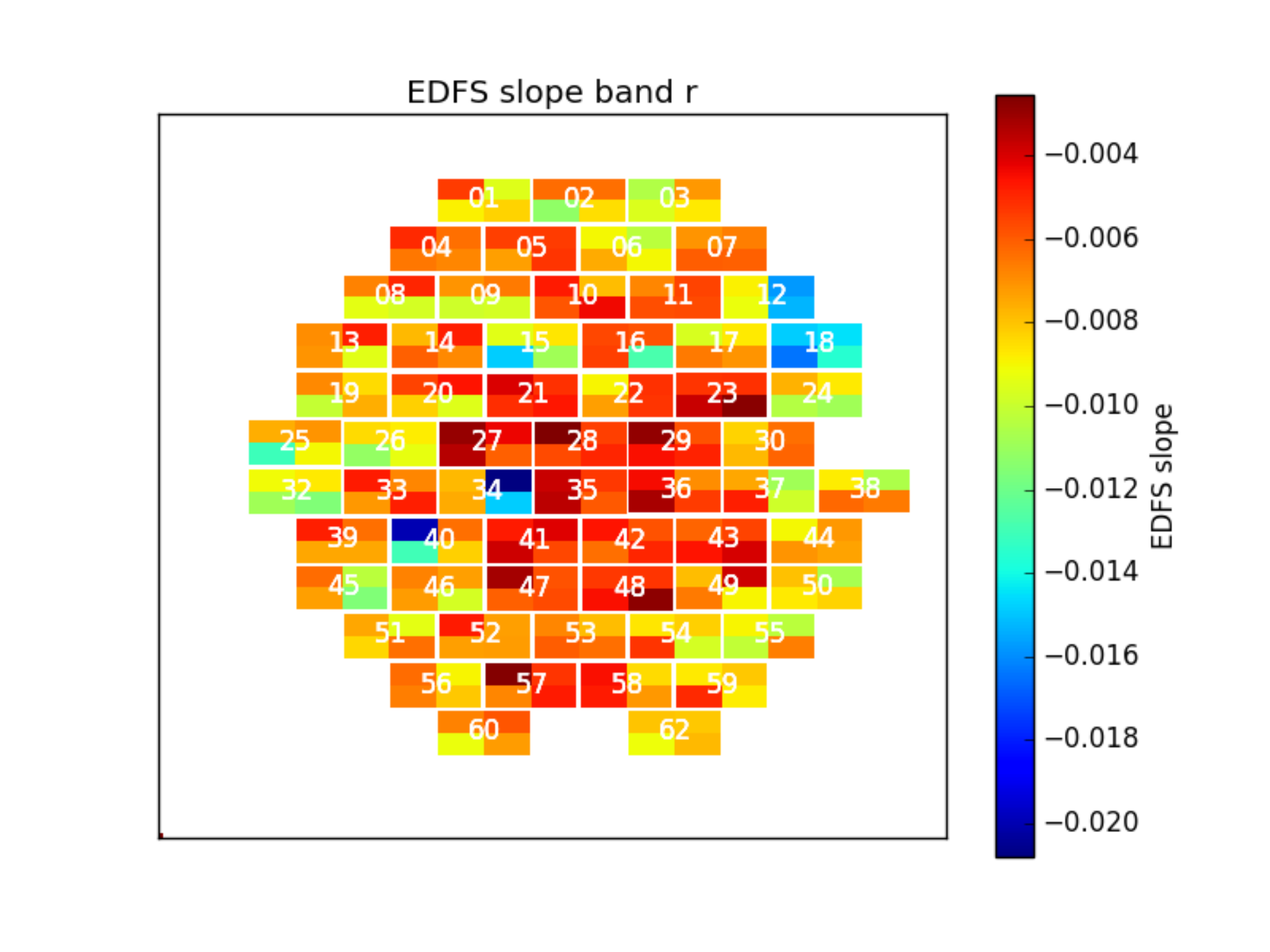}{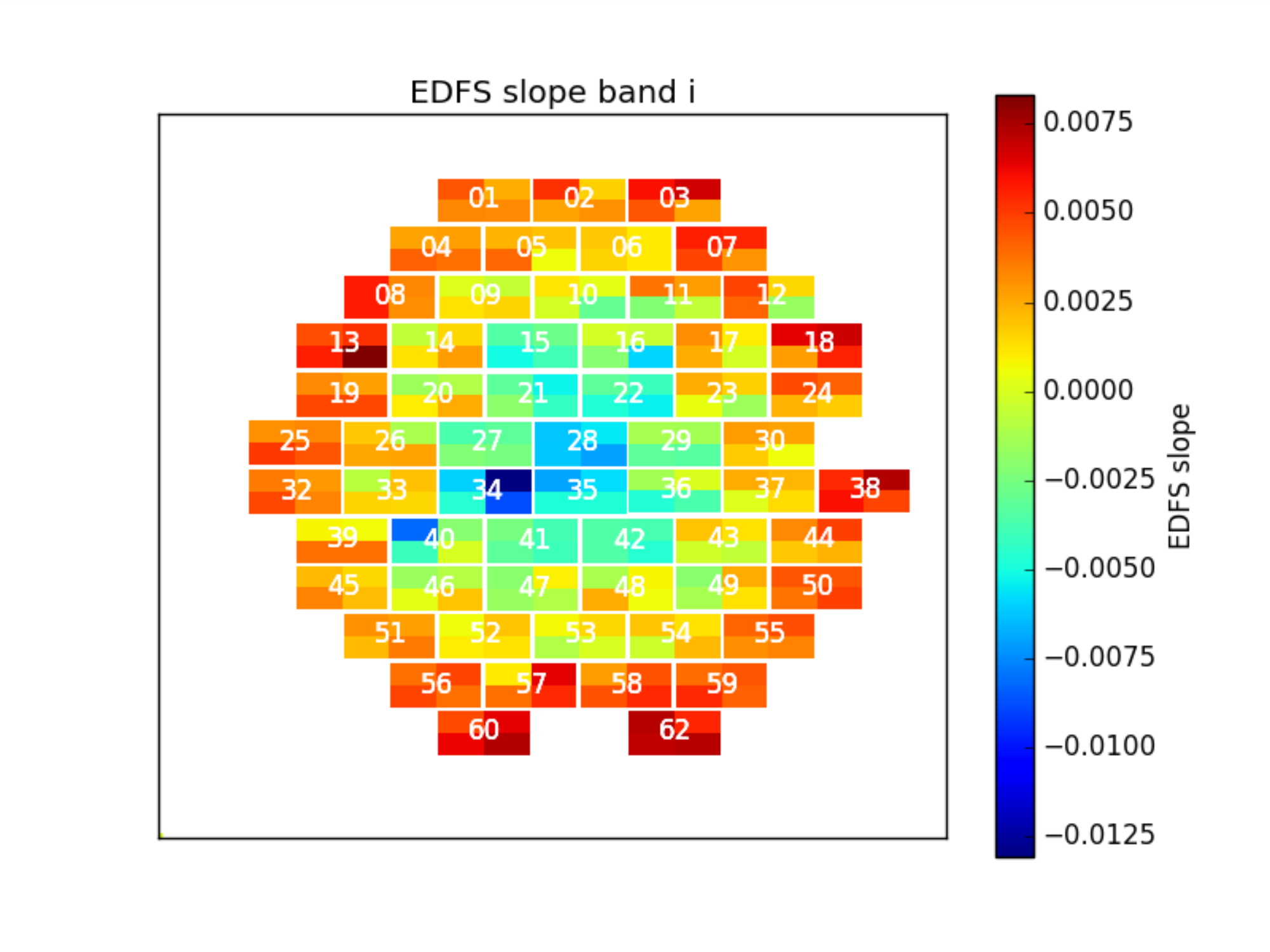}{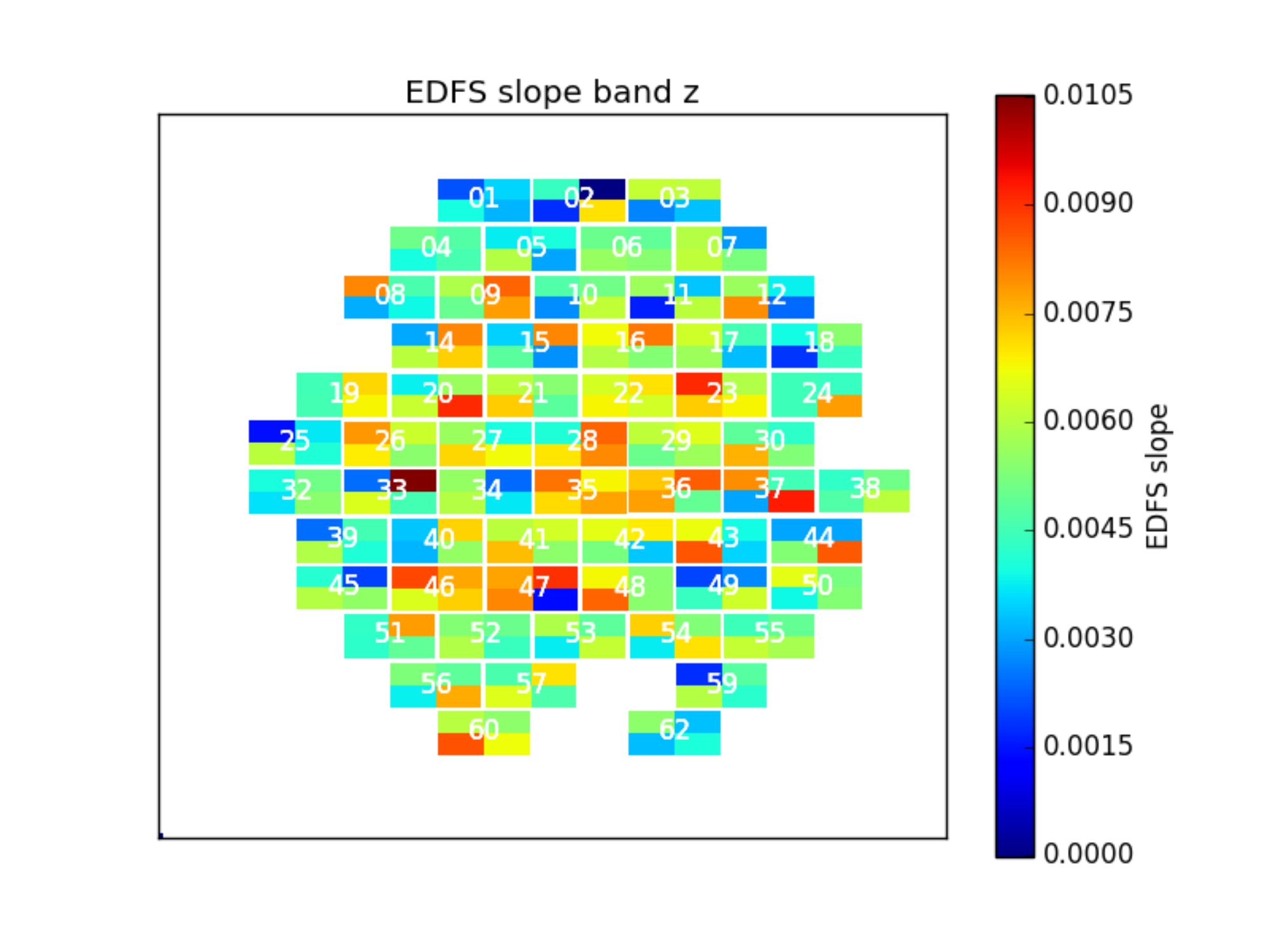}{fig4}{The focal plane maps of color term variation across 62 CCDs for DES $(g/r/i/z)$ filters are shown. Each CCD is divided into four subregions. Variations from zero indicate that the bandpass for the CCD subregion differs from the average DECam bandpass.}


\acknowledgments

The authors acknowledge the Euclid Consortium, the European Space Agency and the support of a number of agencies and institutes that have supported the development of Euclid. A detailed complete list is available on the Euclid web site (http://www.euclid-ec.org). In particular the Academy of Finland, the Agenzia Spaziale Italiana,
the Belgian Science Policy, the Canadian Euclid Consortium, the Centre National d'Etudes Spatiales, the Deutches Zentrum f\"ur Luft- and Raumfahrt, the Danish Space Research Institute, the Funda\c{c}\~{a}o para a Ci\^{e}nca e a Tecnologia, the Ministerio de Economia y Competitividad, the National Aeronautics and Space Administration, the Netherlandse Onderzoekschool Voor Astronomie, the Norvegian Space Center, the Romanian Space Agency, the State Secretariat for Education, Research and Innovation (SERI) at the Swiss Space Office (SSO), and the United Kingdom Space Agency.


\bibliography{P11-5}  


\end{document}